\documentclass[12pt]{iopart}
\usepackage{graphicx}
\usepackage{iopams}  
\usepackage{setstack}

\def\ve{\varepsilon}
\def\dc{\partial}
\def\R{{\mathbb R}}

\begin{document}

\title{Trapped modes in finite quantum waveguides}

\author{Andrey L. Delitsyn$^1$, Binh-Thanh Nguyen$^2$ and Denis~S.~Grebenkov$^{2-4}$}

\address{$^1$
Mathematical Department of the Faculty of Physics, Moscow State University, 119991 Moscow, Russia }
  \ead{delitsyn@mail.ru}

\address{$^2$
Laboratoire de Physique de la Mati\`{e}re Condens\'{e}e (UMR 7643), CNRS -- Ecole Polytechnique, F-91128 Palaiseau, France}
  \ead{binh-thanh.nguyen@polytechnique.edu}

\address{$^3$
Laboratoire Poncelet (UMI 2615), CNRS -- Independent University of Moscow, Bolshoy Vlasyevskiy Pereulok 11, 119002 Moscow, Russia }

\address{$^4$
Chebyshev Laboratory, Saint Petersburg State University, 14th line of Vasil'evskiy Ostrov 29, Saint Petersburg, Russia }
  \ead{denis.grebenkov@polytechnique.edu}

\begin{abstract}
The Laplace operator in infinite quantum waveguides (e.g., a bent
strip or a twisted tube) often has a point-like eigenvalue below the
essential spectrum that corresponds to a trapped eigenmode of finite
L2 norm.  We revisit this statement for resonators with long but
finite branches that we call ``finite waveguides''.  Although now
there is no essential spectrum and all eigenfunctions have finite L2
norm, the trapping can be understood as an exponential decay of the
eigenfunction inside the branches.  We describe a general variational
formalism for detecting trapped modes in such resonators.  For finite
waveguides with general cylindrical branches, we obtain a sufficient
condition which determines the minimal length of branches for getting
a trapped eigenmode.  Varying the branch lengths may switch certain
eigenmodes from non-trapped to trapped states.  These concepts are
illustrated for several typical waveguides (L-shape, bent strip,
crossing of two stripes, etc.).  We conclude that the well-established
theory of trapping in infinite waveguides may be incomplete and
require further development for being applied to microscopic quantum
devices.
\end{abstract}
\pacs{ 02.30.Jr, 41.20.Cv, 41.20.Jb, 03.65.Ge }


\noindent{\it Keywords\/}: eigenfunction, Laplace operator, trapping, localization, decay, resonator


\maketitle

\section{ Introduction }

By its name, a waveguide serves for propagating waves which may be of
different physical origins: fluctuations of pressure in acoustics,
electromagnetic waves in optics, particle waves in quantum mechanics,
surface water waves in hydrodynamics, etc.  The transmission
properties of a waveguide can be characterized by its resonance
frequencies or, equivalently, by the spectrum of an operator which
describes the waves motion (e.g., the Laplace operator in the most
usual case).  For infinite waveguides, the spectrum consists of two
parts: (i) the essential (or continuous) spectrum for which the
related resonances are extended over the whole domain and thus have
infinite $L_2$ norms, and (ii) the discrete (or point-like) spectrum
for which the related eigenfunctions have finite $L_2$ norms and thus
necessarily ``trapped'' or ``localized'' in a region of the waveguide.
A wave excited at the frequency of the trapped eigenmode remains in
the localization region and does not propagate.

The existence of trapped, bound or localized eigenmodes in classical
and quantum waveguides has been thoroughly investigated (see reviews
\cite{Duclos95,Linton07} and also references in \cite{Olendski10}).
In the seminal paper, Rellich proved the existence of a localized
eigenfunction in a deformed infinite cylinder \cite{Rellich}.  His
results were significantly extended by Jones \cite{Jones53}.  Ursell
reported on the existence of trapped modes in surface water waves in
channels \cite{Ursell51,Ursell87,Ursell91}, while Parker observed
experimentally the trapped modes in locally perturbed acoustic
waveguides \cite{Parker66,Parker67}.  Exner and Seba considered an
infinite bent strip of smooth curvature and showed the existence of
trapped modes by reducing the problem to Schr\"odinger operator in the
straight strip, with the potential depending on the curvature
\cite{Exner89}.  Goldstone and Jaffe gave the variational proof that the
wave equation subject to Dirichlet boundary condition always has a
localized eigenmode in an infinite tube of constant cross-section in
any dimension, provided that the tube is not exactly straight
\cite{Goldstone92}.  The problem of localization in acoustic
waveguides with Neumann boundary condition has also been investigated
\cite{Evans92,Evans94}.  For instance, Evans {\it et al.}  considered
a straight strip with an inclusion of arbitrary (but symmetric) shape
\cite{Evans94} (see \cite{Davies98} for further extension).  Such an
inclusion obstructed the propagation of waves and was shown to result
in trapped modes.  The effect of mixed Dirichlet, Neumann and Robin
boundary conditions on the localization was also investigated (see
\cite{Olendski10,Bulla97} and references therein).  A mathematical
analysis of guided water waves was developed in \cite{Bonnet-Ben93}.

All the aforementioned works dealt with infinite waveguides for which
the Laplace operator spectrum is essential, with possible inclusion of
discrete points.  Since these discrete points were responsible for
trapped modes, the major question was whether or not such discrete
points exist below the essential spectrum.  It is worth noting that
the localized modes have to decay relatively fast at infinity in order
to guarantee the finite $L_2$ norm.  But the same question about the
existence of rapidly decaying eigenfunctions may be formulated for
bounded domains (resonators) with long branches that we call ``finite
waveguides'' (Fig. \ref{fig:domain}).  This problem is different in
many aspects.  Since all eigenfunctions have now finite $L_2$ norms,
the definition of trapped or localized modes has to be revised.  Quite
surprisingly, a rigorous definition of localization in bounded domains
turns out to be a challenging problem \cite{Sapoval91,Even99,Felix07}.
In the context of the present paper concerning finite waveguides, an
eigenmode is called trapped or localized if it decays exponentially
fast in prominent subregions (branches) of the bounded domain.  The
exponential decay of an eigenfunction in the branch can be related to
the smallness of the associated eigenvalue in comparison to the
cut-off frequency, i.e. the first eigenvalue of the Laplace operator
in the cross-section of that branch \cite{Jackson}.  In other words,
the existence of a trapped mode is related to ``smallness'' of the
eigenvalue, in full analogy to infinite waveguides.  Using the
standard mathematical tools such as domain decomposition, explicit
representation of solutions of the Helmholtz equation and variational
principle, we aim at formalizing these ideas and providing a
sufficient condition on the branch lengths for getting a trapped mode.
The dependence of the localization character on the length of branches
is the main result of the paper and a new feature of finite waveguides
which was overseen in the well-established theory of infinite
waveguides.  As in practice all quantum waveguides are finite, this
dependence may be important for microelectronic devices.

The paper is organized as follows.  In Sec. \ref{sec:main}, we adapt
the method by Bonnet-Ben Dhia and Joly \cite{Bonnet-Ben93} in order to
reduce the original eigenvalue problem in the whole domain to the
nonlinear eigenvalue problem in the domain without branches.  Although
the new problem is more sophisticated, its variational reformulation
provides a general framework for proving the trapping (or
localization) of eigenfunctions.  We use it to derive the main result
of the paper: a sufficient condition (\ref{eq:cond_general}) on the
branch lengths for getting a trapped mode.  In sharp contrast to an
infinite (non-straight) waveguide of a constant cross-section, for
which the first eigenfunction is always trapped and exponentially
decaying \cite{Goldstone92}, finite waveguides may {\it or may not}
have such an eigenfunction, depending on the length of branches.  This
method is then illustrated in Sec. \ref{sec:examples} for several
finite waveguides (e.g., a bent strip and a cross of two stripes).
For these examples, we estimate the minimal branch length which is
sufficient for getting at least one localized mode.  At the same time,
we provide an example of a waveguide, for which there is no
localization for any branch length.  We also construct a family of
finite waveguides for which the minimal branch length varies
continuously.  As a consequence, for a given (large enough) branch
length, one can construct two almost identical resonators, one with
and the other without localized mode.  This observation may be used
for developing quantum switching devices.

\section{ Theoretical results }
\label{sec:main}

For the sake of clarity, we focus on planar bounded domains with
rectangular branches, while the extension to arbitrary domains in
$\R^n$ with general cylindrical branches is straightforward and
provided at the end of this Section.

\begin{figure}
\begin{center}
\includegraphics[width=120mm]{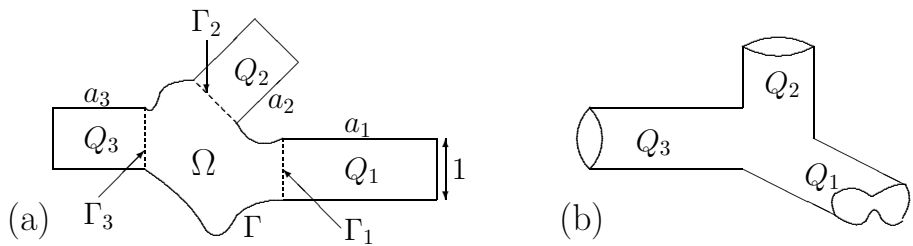}
\end{center}
\caption{
Two examples of a finite waveguide: (a) a planar bounded domain $D$
which is composed of a basic domain $\Omega$ of arbitrary shape and
three rectangular branches $Q_i$ of lengths $a_i$ and width $b=1$; (b)
a three-dimensional bounded domain with three general cylindrical
branches. }
\label{fig:domain}
\end{figure}

We consider a planar bounded domain $D$ composed of a basic domain
$\Omega$ of arbitrary shape and $M$ rectangular branches $Q_i$ of
lengths $a_i$ and width $b$ as shown on Fig. \ref{fig:domain}:
\begin{equation*}
D = \Omega\cup \bigcup\limits_{i=1}^M Q_i .
\end{equation*}
We denote $\Gamma_i = \dc \Omega \cap \dc Q_i$ the inner boundary
between the basic domain $\Omega$ and the branch $Q_i$ and $\Gamma =
\dc\Omega \backslash \bigcup_{i=1}^M \Gamma_i$ the exterior boundary
of $\Omega$.  We study the eigenvalue problem for the Laplace operator
with Dirichlet boundary condition
\begin{equation}
 \label{1}
-\Delta U = \lambda U , \quad U|_{\partial D}=0 .
\end{equation}

\subsection{ Solution in rectangular branches }

Let $u_i(x,y)$ denote the restriction of the solution $U(x,y)$ of the
eigenvalue problem (\ref{1}) to the branch $Q_i$.  For convenience, we
take $b=1$ and assume that the coordinates $x$ and $y$ are chosen in
such a way that $Q_i = \{ (x,y)\in\R^2~:~ 0<x<a_i,~ 0<y<1 \}$ (the
final result will not depend on this particular coordinate system).
The eigenfunction $u_i(x,y)$ satisfying Dirichlet boundary condition
on $\partial Q_i$ has the standard representation:
\begin{equation}
\label{2}
u_i(x,y) \equiv U_{|Q_i} = \sum\limits_{n=1}^\infty c_n \sinh(\gamma_n (a_i-x)) \sin (\pi n y) , 
\end{equation}
where $\gamma_n = \sqrt{\pi^2 n^2 - \lambda}$ and $c_n$ are the
Fourier coefficients of the function $U$ at the inner boundary
$\Gamma_i$ (at $x = 0$):
\begin{equation}
\label{eq:cn}
c_n = \frac{2}{\sinh(\gamma_n a_i)} \int\limits_0^1 dy ~ U(0,y) \sin (\pi n y) .
\end{equation}
Substituting this relation into Eq. (\ref{2}) yields
\begin{equation}
\label{eq:u1u2}
u_i(x,y) = 2\sum\limits_{n=1}^{\infty} \bigl(U_{|\Gamma_i}, \sin (\pi n y)\bigr)_{L_2(\Gamma_i)}  
\frac{ \sinh (\gamma_n(a_i - x))}{\sinh (\gamma_n a_i)} \sin (\pi n y) ,
\end{equation}
where the integral in (\ref{eq:cn}) was interpreted as the scalar
product in $L_2(\Gamma_i)$.  The representation (\ref{eq:u1u2}) is of
course formal because its coefficients are still unknown.
Nevertheless, one can already distinguish two different cases.

(i) If $\lambda < \pi^2$, all $\gamma_n$ are real, and the
representation (\ref{eq:u1u2}) decays exponentially.  
In fact, writing the squared $L_2$-norm of the function $u_i(x,y)$
along the vertical cross-section of the branch $Q_i$ at $x$,
\begin{equation*}
I_i(x) \equiv \int\limits_0^1 u_i^2(x,y) dy 
= 2\sum \limits_{n=1}^{\infty} \bigl(U_{|\Gamma_i}, \sin (\pi n y)\bigr)^2_{L_2(\Gamma_i)}
\frac{\sinh^2(\gamma_n(a_i - x))}{\sinh^2(\gamma_n a_i)} ,
\end{equation*}
one can use the inequality $\sinh(\gamma_n (a_i-x)) \leq
\sinh(\gamma_n a_i) e^{-\gamma_1 x}$ to get
\begin{equation}
\label{eq:decay}
I_i(x) \leq 2 \sum\limits_{n=1}^{\infty} \bigl(U_{|\Gamma_i}, \sin(\pi n y)\bigr)^2_{L_2(\Gamma_i)} e^{-2 \gamma_1 x}
= I_i(0) e^{-2\gamma_1 x}  \quad (0 < x < a_i).
\end{equation}
This shows the exponential decay along the branch with the decay rate
$2\gamma_1 = 2\sqrt{\pi^2 - \lambda}$. 

(ii) In turn, if $\lambda > \pi^2$, some $\gamma_n$ are purely
imaginary so that $\sinh(\gamma_n z)$ becomes $\sin(|\gamma_n|z)$, and
the exponential decay is replaced by an oscillating behavior.

One sees that the problem of localization of the eigenfunction in the
basic domain $\Omega$ is reduced to checking whether the eigenvalue
$\lambda$ is smaller or greater than $\pi^2$ (or $\pi^2/b^2$ in
general).

\subsection{ Nonlinear eigenvalue problem }

The explicit representation (\ref{eq:u1u2}) of the eigenfunction in
the branch $Q_i$ allows one to reformulate the original eigenvalue
problem (\ref{1}) in the whole domain $D$ as a specific eigenvalue
problem in the basic domain $\Omega$.  In fact, the restriction of $U$
onto the basic domain $\Omega$, $u \equiv U_{|\Omega}$, satisfies the
following equations
\begin{equation}
\label{eq:eigen_basic0}
-\Delta u = \lambda u  \quad \mathrm{in}~\Omega, \quad u|_{\Gamma}=0 , \quad   u|_{\Gamma_i} = u_i|_{\Gamma_i},
\quad  \frac{\dc u}{\dc n}|_{\Gamma_i} = - \frac{\dc u_i}{\dc n}|_{\Gamma_i} ,
\end{equation}
where $\dc/\dc n$ denotes the normal derivative directed outwards the
domain.  The last two conditions ensure that the eigenfunction and its
derivative are continuous at inner boundaries $\Gamma_i$ (the sign
minus accounting for opposite orientations of the normal derivatives
on both sides of the inner boundary).  The normal derivative of $u_i$
can be explicitly written by using Eq. (\ref{eq:u1u2}):
\begin{equation}
\label{eq:aux1}
\frac{\dc u_i}{\dc n}|_{\Gamma_i} = - \frac{\dc u_i}{\dc x}|_{x=0} =
2\sum\limits_{n=1}^{\infty} \gamma_n \coth(\gamma_n a_i) \bigl(U_{|\Gamma_i}, \sin (\pi n y)\bigr)_{L_2(\Gamma_i)} \sin (\pi n y) .
\end{equation}
Denoting $T_i(\lambda)$ an operator acting from $H^{1/2}(\Gamma_i)$ to
$H^{-1/2}(\Gamma_i)$ (see \cite{Lions} for details) as
\begin{equation*}
T_i(\lambda) f \equiv 2\sum\limits_{n=1}^{\infty} \gamma_n \coth(\gamma_n a_i) \bigl(f, \sin (\pi n y)\bigr)_{L_2(\Gamma_i)} \sin (\pi n y) ,
\end{equation*}
the right-hand side of Eq. (\ref{eq:aux1}) can be written as
\begin{equation*}
\frac{\dc u_i}{\dc n}|_{\Gamma_i} = T_i(\lambda) U_{|\Gamma_i} .
\end{equation*}
The eigenvalue problem (\ref{eq:eigen_basic0}) admits thus a closed
representation as
\begin{equation}
\label{eq:eigen_basic}
-\Delta u = \lambda u  \quad \mathrm{in}~\Omega, \quad u|_{\Gamma}=0 , \quad   
\frac{\dc u}{\dc n}|_{\Gamma_i} = - T_i(\lambda) u|_{\Gamma_i} .
\end{equation}
The presence of branches and their shapes are fully accounted for by
the operators $T_i$ which are somewhat analogous to
Dirichlet-to-Neumann operators.

Although this domain decomposition allows one to remove the branches
and get a closed formulation for the basic domain $\Omega$, the new
eigenvalue problem is {\it nonlinear} because the eigenvalue $\lambda$
appears also in the boundary condition through the operators
$T_i(\lambda)$.  A trick to overcome this difficulty goes back to the
Birman-Schwinger method \cite{Birman61,Schwinger61} (see also
\cite{Aslanyan81}).  Following \cite{Bonnet-Ben93}, we fix $\lambda$
and solve the {\it linear} eigenvalue problem
\begin{equation}
\label{eq:eigen_basic2}
-\Delta u = \mu(\lambda) u  \quad \mathrm{in}~\Omega, \quad u|_{\Gamma}=0 , \quad   
\frac{\dc u}{\dc n}|_{\Gamma_i} = - T_i(\lambda) u|_{\Gamma_i} ,
\end{equation}
where $\mu(\lambda)$ denotes the eigenvalue which is parameterized by
$\lambda$.  The solution of the original problem is recovered when
$\mu(\lambda) = \lambda$.

From a practical point of view, a numerical solution of
Eqs. (\ref{eq:eigen_basic2}) with the subsequent resolution of the
equation $\mu(\lambda)=\lambda$ is in general much more difficult than
solving the original eigenvalue problem (see also \cite{Levitin08} for
possible numerical schemes).  In turn, Eqs. (\ref{eq:eigen_basic2})
are convenient for checking whether the first eigenvalue $\lambda_1$
is smaller or greater than $\pi^2$, as explained below.

\subsection{ Variational formulation }

We search for a weak solution of the eigenvalue problem
(\ref{eq:eigen_basic2}) in the Sobolev space
\begin{equation*}
H^1_0(\Omega) = \{ v(x,y) \in L_2(\Omega), ~ \dc v/\dc x \in L_2(\Omega),~ \dc v/\dc y \in L_2(\Omega),~ v|_{\Gamma} = 0\} .
\end{equation*}
Multiplying Eq. (\ref{eq:eigen_basic2}) by a trial function $v\in
H^1_0(\Omega)$ and integrating by parts, one gets
\begin{equation*}
\mu(\lambda) \int\limits_{\Omega} v u = - \int\limits_{\Omega} v \Delta u = \int\limits_{\Omega} (\nabla v, \nabla u) 
- \int\limits_{\partial \Omega} v \frac{\partial u}{\partial n} .
\end{equation*}
Since $v$ vanishes on $\Gamma$, the weak formulation of the problem
reads as
\begin{equation}
(\nabla u,\nabla v)_{L_2(\Omega)} + \sum\limits_{i=1}^M \bigl(T_i(\lambda) u, v\bigr)_{L_2(\Gamma_i)} 
= \mu(\lambda) \bigl(u,v\bigr)_{L_2(\Omega)}~~ \forall v \in H^1_0(\Omega) .
\end{equation}
The first eigenvalue $\mu_1(\lambda)$ is then obtained from the
Rayleigh's principle
\begin{equation}
\label{eq:nu1}
\mu_1(\lambda) = \inf\limits_{v\in H^1_0(\Omega),~ v\ne 0} \frac{\bigl(\nabla v,\nabla v\bigr)_{L_2(\Omega)}+
\sum\nolimits_{i=1}^M \bigl(T_i(\lambda) v, v\bigr)_{L_2(\Gamma_i)}}{\bigl(v,v\bigr)_{L_2(\Omega)}} .
\end{equation}

One can show that $\mu_1(\lambda)$ is a continuous monotonously
decreasing function of $\lambda$ on the segment $(0, \pi^2]$.  For
this purpose, one first computes explicitly the derivative of the
function
\begin{equation*}
h(\lambda) \equiv \gamma_n \coth(\gamma_n a_i) = \sqrt{\pi^2 n^2 - \lambda} \coth(\sqrt{\pi^2 n^2 - \lambda}~ a_i)
\end{equation*}
and checks that $h'(\lambda) < 0$.  Now one can show that
$\mu_1(\lambda_1) \leq \mu_1(\lambda_2)$ if $\lambda_1 > \lambda_2$.
If some trial function $v_2$ minimizes the Rayleigh's quotient
(\ref{eq:nu1}) for $\lambda_2$, one has
\begin{eqnarray*}
\mu_1(\lambda_1) & \leq & \frac{\bigl(\nabla v_2,\nabla v_2\bigr)_{L_2(\Omega)}+
\sum\nolimits_{i=1}^M \bigl(T_i(\lambda_1) v_2, v_2\bigr)_{L_2(\Gamma_i)}}{\bigl(v_2,v_2\bigr)_{L_2(\Omega)}} \\
& \leq & \frac{\bigl(\nabla v_2,\nabla v_2\bigr)_{L_2(\Omega)}+
\sum\nolimits_{i=1}^M \bigl(T_i(\lambda_2) v_2, v_2\bigr)_{L_2(\Gamma_i)}}{\bigl(v_2,v_2\bigr)_{L_2(\Omega)}} = \mu_1(\lambda_2) ,
\end{eqnarray*}
where the monotonous decrease of the function $h(\lambda)$ was used
(the mathematical proof of the continuity for an analogous functional
is given in \cite{Delitsyn04}).

Since the function $\mu_1(\lambda)$ is positive, continuous and
monotonously decreasing, the equation $\mu_1(\lambda) = \lambda$ has a
solution $0 < \lambda < \pi^2$ if and only if $\mu_1(\pi^2) < \pi^2$.
This is a necessary and sufficient condition for getting a trapped
mode for the linear eigenvalue problem (\ref{eq:eigen_basic2}).

\subsection{ Sufficient condition }

For any trial function $v\in H^1_0(\Omega)$, we denote the Rayleigh's
quotient
\begin{equation}
\mu(v) = \frac{\bigl(\nabla v,\nabla v\bigr)_{L_2(\Omega)} +
\sum\nolimits_{i=1}^M \bigl(T_i(\pi^2) v, v\bigr)_{L_2(\Gamma_i)}}{\bigl(v,v\bigr)_{L_2(\Omega)}} .
\end{equation}
Since $\gamma_n(\pi^2) = \pi \sqrt{n^2-1}$, one has $\gamma_1(\pi^2) =
0$, and the operators $T_i(\pi^2)$ can be decomposed into two parts so
that
\begin{eqnarray}
\mu(v) &=& \bigl(v,v\bigr)^{-1}_{L_2(\Omega)} \biggl\{\bigl(\nabla v,\nabla v\bigr)_{L_2(\Omega)} +
2\sum\limits_{i=1}^M \frac{1}{a_i} \bigl(v, \sin(\pi y)\bigr)^2_{L_2(\Gamma_i)}  \nonumber \\
& & + 2\pi \sum\limits_{n=2}^\infty \sqrt{n^2-1} \sum\limits_{i=1}^M \coth(\pi a_i\sqrt{n^2-1}) 
\bigl(v, \sin(\pi ny)\bigr)^2_{L_2(\Gamma_i)}\biggr\} . 
\end{eqnarray}
If one finds a trial function $v\in H^1_0(\Omega)$ for which $\mu(v) <
\pi^2$, then the first eigenvalue $\mu_1(\pi^2)$ necessarily satisfies
this condition because $\mu_1(\pi^2) \leq \mu(v)$.  The inequality
$\mu(v) <\pi^2$ is thus a sufficient (but not necessary) condition.
Given that $\coth(\pi a_i \sqrt{n^2-1}) \leq \coth(\pi a_i\sqrt{3})$
for any $n\geq 2$, the sufficient condition can be written as
\begin{equation}
\label{eq:cond}
\sum\limits_{i=1}^M \frac{\sigma_i}{a_i} < \beta - \sum\limits_{i=1}^M \kappa_i \coth(\pi a_i\sqrt{3}) ,
\end{equation}
where
\begin{eqnarray}
\beta    &=& \pi^2 \bigl(v,v\bigr)_{L_2(\Omega)} - \bigl(\nabla v,\nabla v\bigr)_{L_2(\Omega)} , \\
\sigma_i &=& 2\bigl(v, \sin(\pi y)\bigr)^2_{L_2(\Gamma_i)} , \\
\label{eq:kappa_i}
\kappa_i &=& 2\pi \sum\limits_{n=2}^\infty \sqrt{n^2-1} ~\bigl(v, \sin(\pi n y)\bigr)^2_{L_2(\Gamma_i)} . 
\end{eqnarray}

Before moving to examples, several remarks are in order.

(i) The shape of the branches enters through the operator
$T_i(\lambda)$.  Although the above analysis was presented for
rectangular branches, its extension to bounded domains in $\R^n$ with
general cylindrical branches is straightforward and based on the
variable separation (in directions parallel and perpendicular to the
branch).  In fact, the Fourier coefficients $(u, \sin(\pi
ny))_{L_2(\Gamma_i)}$ on the unit interval (i.e., the cross-section of
the rectangular branch) have to be replaced by a spectral
decomposition over the orthonormal eigenfunctions
$\{\psi_n(y)\}_{n=1}^\infty$ of the Laplace operator $\Delta_\perp$ in
the cross-section $\Gamma_i$ of the studied branch (in general,
$\Gamma_i$ is a bounded domain in $\R^{n-1}$):
\begin{equation}
\label{eq:psi}
\Delta_\perp \psi_n + \nu_n \psi_n = 0  \quad \mathrm{in}~\Gamma_i, \qquad \psi_n|_{\partial \Gamma_i} = 0.
\end{equation}
In particular, the operator $T_i(\lambda)$ becomes
\begin{equation*}
T_i(\lambda) f = \sum\limits_{n=1}^\infty \gamma_n \coth(\gamma_n a_i) \bigl(f, \psi_n)_{L_2(\Gamma_i)} \psi_n(y) ,
\end{equation*}
with $\gamma_n = \sqrt{\nu_n - \lambda}$.  Repeating the above
analysis, one immediately deduces a sufficient condition for getting a
trapped mode:
\begin{equation}
\label{eq:cond_general}
\sum\limits_{i=1}^M \frac{\sigma_i}{a_i} < \beta - \sum\limits_{i=1}^M \kappa_i \coth(a_i\sqrt{\nu_2 - \nu_1}) ,
\end{equation}
with
\begin{eqnarray}
\label{eq:beta_gen}
\beta    &=& \nu_1 \bigl(v,v\bigr)_{L_2(\Omega)} - \bigl(\nabla v,\nabla v\bigr)_{L_2(\Omega)} , \\
\label{eq:sigma_gen}
\sigma_i &=& \bigl(v, \psi_1\bigr)^2_{L_2(\Gamma_i)} , \\
\label{eq:kappa_gen}
\kappa_i &=& \sum\limits_{n=2}^\infty \sqrt{\nu_n-\nu_1} ~\bigl(v, \psi_n\bigr)^2_{L_2(\Gamma_i)} . 
\end{eqnarray}
One retrieves the above results for rectangular branches by putting
$\psi_n(y) = \sqrt{2}\sin(\pi n y)$ and $\nu_n = \pi^2 n^2$.

The inequality (\ref{eq:cond_general}) is the main result of the
paper.  Although there is no explicit recipe for choosing the trial
function $v$ (which determines the coefficients $\beta$, $\sigma_i$
and $\kappa_i$), this is a general framework for studying the
localization (or trapping) in domains with cylindrical branches.

(ii) If the branches are long enough (e.g., $a_i \gg
(\nu_2-\nu_1)^{-1/2}$), the values $\coth (a_i \sqrt{\nu_2-\nu_1})$
are very close to $1$ and can be replaced by $1+\epsilon$ where
$\epsilon$ is set by the expected minimal length so that the
inequality (\ref{eq:cond_general}) becomes more explicit in terms of
$a_i$:
\begin{equation}
\label{eq:cond2}
\sum\limits_{i=1}^M \frac{\sigma_i}{a_i} < \beta - (1+\epsilon)\sum\limits_{i=1}^M \kappa_i .
\end{equation}
In the particular case when all $\sigma_i$ are the same, one can
introduce the threshold value $\eta$ as
\begin{equation}
\label{eq:eta}
\sum\limits_{i=1}^M \frac{1}{a_i} < \eta ,  \qquad \eta \equiv \frac{\beta}{\sigma_1} - 
\frac{(1+\epsilon)}{\sigma_1}\sum\limits_{i=1}^M \kappa_i .
\end{equation}
For domains with identical branches, $a_i = a$, the above condition
determines the branch length $a_{\rm th} = M/\eta$ which is long
enough for the emergence of localization.  This means that for any $a
> a_{\rm th}$ there is a localized eigenmode.  Since $a_{\rm th}$ was
obtained from the sufficient condition (\ref{eq:cond_general}), the
opposite statement is not true: for $a < a_{\rm th}$, this condition
does not indicate whether the eigenfunction is localized or not.  In
fact, $a_{\rm th}$ is the upper bound for the minimal branch length
$a_{\rm min}$ which may distinguish waveguides with and without
localized modes (see Sec. \ref{sec:examples}).

(iii) The trial function should be chosen to ensure the convergence of
the series in (\ref{eq:kappa_gen}).  If the boundary of $\Omega$ is
smooth, the series in (\ref{eq:kappa_gen}) converges for any function
$v$ from $H^1_0(\Omega)$ according to the trace theorem \cite{Lions}.
In turn, the presence of corners or other singularities may require
additional verifications for the convergence, as illustrated in
Sec. \ref{sec:strip}.

(iv) The implementation of various widths $b_i$ of the rectangular
branches is relatively straightforward (e.g., $\sin(\pi ny)$ is
replaced by $\sin(\pi ny/b_i)$, etc.).  In order to guarantee the
exponential decay in all branches, one needs $\lambda < \pi^2/b_i^2$
for all $i$, i.e., $\lambda < \pi^2/\max \{b_i^2\}$.  Rescaling the
whole domain in such a way that $\max \{b_i\} = 1$, one can use the
above conditions.

(v) According to Eq. (\ref{eq:decay}), the decay rate, $2\gamma_1$, is
determined by the eigenvalue $\lambda$.  Since $\mu(v)$ is an upper
bound for the first eigenvalue, one gets a lower bound for the decay
rate:
\begin{equation*}
2\gamma_1 \geq 2\sqrt{\nu_1 - \mu(v)} = 2(v,v)^{-1/2}_{L_2(\Omega)} 
\left(\beta - \sum\limits_{i=1}^M [\sigma_i/a_i + \kappa_i \coth(a_i\sqrt{\nu_2-\nu_1})]\right)^{1/2} .
\end{equation*}

\section{ Several examples }
\label{sec:examples}

As we already mentionned, there is no general recipe for choosing a
trial function $v$ from $H^1_0(\Omega)$.  Of course, the best possible
choice is the eigenfunction on which $\mu(v)$ reaches its minimum.
Except for few cases, the eigenfunction is not known but one can often
guess how it behaves for a given basic domain.  Since the gradient of
the trial function $v$ enters into the coefficient $\beta$ with the
sign minus, slowly varying functions are preferred.  In what follows,
we illustrate these concepts for several examples.

\begin{figure}
\begin{center}
\includegraphics[width=150mm]{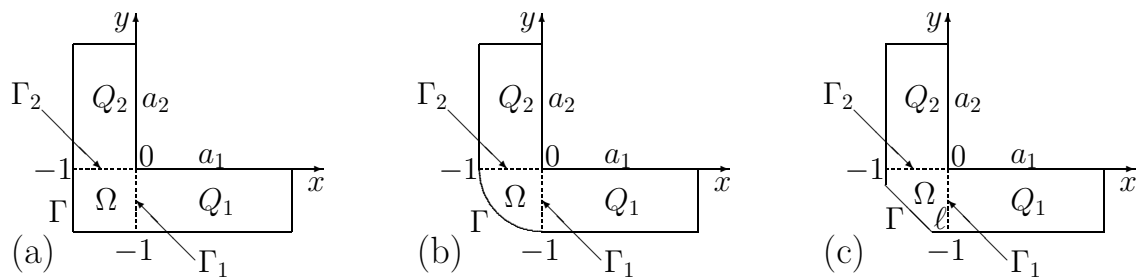}
\end{center}
\caption{
Three types of a bent waveguide: (a) L-shape, (b) bent strip, and (c)
truncated L-shapes parameterized by the length $\ell$ varying from $0$
(triangular basic domain) to $1$ (the original L-shape). }
\label{fig:bent}
\end{figure}

\subsection{ L-shape }

We start by a classical problem of localization in L-shape with two
rectangular branches of lengths $a_1$ and $a_2$ (Fig. \ref{fig:bent}a)
for which the basic domain is simply the unit square.  In the limit
case $a_1 = a_2 = 0$ (i.e., $D = \Omega$, without branches), the first
eigenvalue $\lambda_1 = 2\pi^2 > \pi^2$ so that, according to our
terminology, there is no localization.  Since $\lambda_1$ continuously
varies with $a$ ($a_1 = a_2 = a$), the inequality $\lambda_1 > \pi^2$
also remains true for relatively short branches.  In turn, given that
$\lambda_1 < \pi^2$ for infinitely long branches, there should exist
the minimal branch length $a_{\rm min}$ such that $\lambda_1 = \pi^2$.
At this length, the first eigenfunction passes from non-localized
state ($a < a_{\rm min}$) to localized state ($a > a_{\rm min}$).  In
what follows, we employ the sufficient condition (\ref{eq:cond}) in
order to get the upper bound for $a_{\rm min}$.

The most intuitive choice for the trial function would be the first
eigenfunction for the unit square with Dirichlet boundary condition,
$v(x,y) = \sin(\pi x)\sin(\pi y)$.  However, one easily checks that
$\beta = 0$ for this function, while $\sigma_i$ and $\kappa_i$ are
always non-negative.  As a consequence, the condition (\ref{eq:cond})
is never satisfied for this trial function.  It simply means that the
first choice was wrong.

For the trial function
\begin{equation}
\label{eq:trial_L-shape}
v(x,y) = (1+x) \sin(\pi y) + (1+y) \sin(\pi x) ,
\end{equation}
the explicit integration yields
\begin{eqnarray*}
&& \bigl(v,v\bigr)_{L_2(\Omega)} = \int\limits_{-1}^0 dx \int\limits_{-1}^0 dy \bigl[(1+x) \sin(\pi y) 
+ (1+y) \sin(\pi x)\bigr]^2 = \frac{1}{3} + \frac{2}{\pi^2} , \\
&& \bigl(\nabla v,\nabla v\bigr)_{L_2(\Omega)} = \int\limits_{-1}^0 dx \int\limits_{-1}^0 dy 
 \biggl[\biggl(\frac{\partial v}{\partial x}\biggr)^2 + \biggl(\frac{\partial v}{\partial y}\biggr)^2\biggr] = \frac{\pi^2}{3} + 1 , \\
&& \bigl(v, \sin(\pi y)\bigr)_{L_2(\Gamma_1)} = \int\limits_{-1}^0 dy \bigl[(1+0) \sin(\pi y) + (1+y) \sin(\pi 0)\bigr] \sin (\pi y) = \frac12 , \\
&& \bigl(v, \sin(\pi n y)\bigr)_{L_2(\Gamma_1)} = \bigl(v, \sin(\pi n x)\bigr)_{L_2(\Gamma_2)} = 0 \qquad (n\geq 2) .
\end{eqnarray*}
from which
\begin{equation*}
\beta = 1, \quad   \sigma_1 = \sigma_2 = \frac12,  \quad   \kappa_1 = \kappa_2 = 0 .
\end{equation*}
The condition (\ref{eq:cond}) reads as
\begin{equation}
\label{eq:cond_L-shape}
\frac{1}{a_1} + \frac{1}{a_2} < 2 .
\end{equation}
If the branches have the same length, $a_1 = a_2 = a$, then the upper
bound of the minimal branch length for getting a localized
eigenfunction is given by $a_{\rm th} = 1$.

\begin{figure}
\begin{center}
\includegraphics[width=120mm]{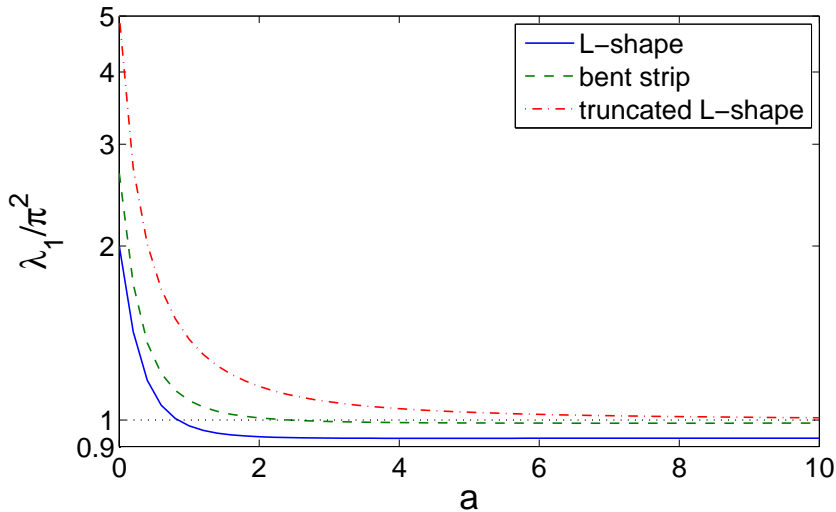}
\end{center}
\caption{
The first eigenvalue $\lambda_1$ divided by $\pi^2$, as a function of
the branch length $a$ ($a_1 = a_2 = a$), for three bent waveguides
shown on Fig. \ref{fig:bent}: L-shape (solid line), bent strip (dashed
line) and truncated L-shape with $\ell = 0$ (dash-dotted line).  For
the first two cases, the curves cross the level $1$ at $a_{\rm
min}\approx 0.84$ and $a_{\rm min}\approx 2.44$, respectively.  In
turn, the third curve always remains greater than $1$ (see
explanations in Sec. \ref{sec:triangular}).  For $a = 0$, $\lambda_1$
is respectively equal to $2\pi^2$, $4\alpha^2$ and $5\pi^2$, where
$\alpha \approx 2.4048$ is the first positive zero of the Bessel
function $J_0(z)$. }
\label{fig:lambda}
\end{figure}

We also solved the original eigenvalue problem (\ref{1}) for L-shape
with $a_1 = a_2 = a$ by a finite element method (FEM) implemented in
Matlab PDEtools.  The first eigenvalue $\lambda_1$ as a function of
the branch length $a$ is shown by solid line on Fig. \ref{fig:lambda}.
One can clearly see a transition from non-localized ($\lambda_1 >
\pi^2$) to localized ($\lambda_1 < \pi^2$) states when $a$ crosses
the minimal branch length $a_{\rm min} \approx 0.84$.  As expected,
the theoretical upper bound $a_{\rm th}$ which was obtained from a
{\it sufficient} condition, exceeds the numerical value $a_{\rm min}$.
In order to improve the theoretical estimate, one has to search for
trial functions which are closer to the true eigenfunction.  At the
same time, $a_{\rm th}$ and $a_{\rm min}$ are close to each other, and
the accuracy of the theoretical result is judged as good.  Similar
results for L-shape in three dimensions are derived in
\ref{sec:Lshape_3D}.

\subsection{ Cross }

\begin{figure}
\begin{center}
\includegraphics[width=140mm]{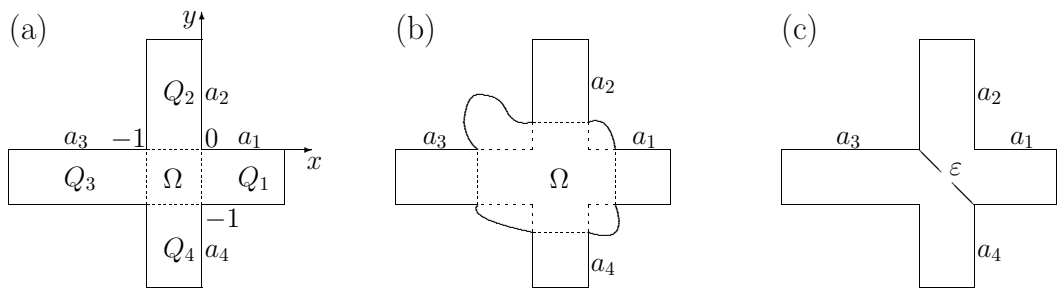}
\end{center}
\caption{
(a) Crossing of two rectangular branches; (b) an extension of the
related basic domain $\Omega$; and (c) coupling between two waveguides
from Fig. \ref{fig:bent}c ($\ell = 0$) with an opening of size
$\ve$. }
\label{fig:cross}
\end{figure}

Another example is a crossing of two perpendicular rectangular
branches (Fig. \ref{fig:cross}a), for which the basic domain is again
the unit square.  Since the trial function (\ref{eq:trial_L-shape})
also satifies the boundary condition for this problem, the previous
sufficient condition (\ref{eq:cond_L-shape}) remains applicable for
arbitrary lengths $a_3$ and $a_4$.  This is not surprising because any
increase of the basic domain (i.e., if the basic domain was considered
as the unit square with two branches $Q_3$ and $Q_4$) decreases the
eigenvalue.  A symmetry argument implies that other consecutive pairs
of branch lengths can be used in the condition
(\ref{eq:cond_L-shape}), e.g., the eigenfunction is localized if
$1/a_2 + 1/a_3 < 2$ for arbitrary $a_1$ and $a_4$.  In turn, the
condition $1/a_1 + 1/a_3 < 2$ is not sufficient for localization (in
fact, taking $a_2 = a_4 = 0$ yields a rectangle without localization).

The specific feature of the cross is that the exterior boundary of the
basic domain $\Omega$ consists of 4 corner points.  We suggest another
trial function
\begin{equation}
v(x,y) = x(1+x) + y(1+y) ,
\end{equation}
which satisfies the Dirichlet boundary condition at these points.  The
direct integration yields
\begin{eqnarray*}
\beta &=& \frac{11}{90}\pi^2 - \frac{2}{3}, \qquad   \sigma_i = \frac{64}{\pi^6} , \\
\kappa_i &=& 2\pi \sum\limits_{n=2}^\infty \sqrt{n^2-1} \left(2\frac{1-(-1)^n}{\pi^3 n^3}\right)^2 \approx 0.0029 .
\end{eqnarray*}
The condition (\ref{eq:cond}) reads now as
\begin{equation}
\label{eq:cross_cond}
\sum\limits_{i=1}^4 \frac{1}{a_i} < \frac{\beta}{\sigma_1} - \frac{\kappa_1}{\sigma_1} \sum\limits_{i=1}^4 \coth(\pi a_i \sqrt{3}) .
\end{equation}
If all the branches have the same length $a$, the upper bound of the
minimal branch length can be estimated by solving the equation
\begin{equation*}
\frac{4}{a_{\rm th}} = \frac{\beta}{\sigma_1} - \frac{4\kappa_1}{\sigma_1} \coth(\pi a_{\rm th}\sqrt{3}) ,
\end{equation*}
from which one gets $a_{\rm th}\approx 0.2407$.  Note that this result
proves and further extends the prediction of localized eigenmodes in
the crossing of infinite rectangular stripes which was made by Schult
{\it et al.} by numerical computation \cite{Schult89}.  In that
reference, the importance of localized electron eigenstates in four
terminal junctions of quantum wires was discussed.

\begin{figure}
\begin{center}
\includegraphics[width=120mm]{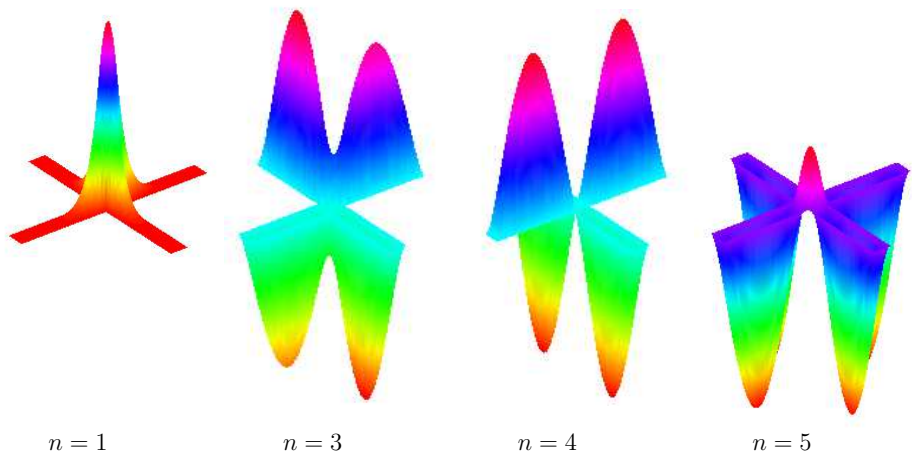}
\end{center}
\caption{
First eigenfunctions for the crossing of two rectangular branches
($a_i = 5$).  The associated eigenvalues are $\lambda_1 \approx
0.661\pi^2$, $\lambda_2 = \lambda_3 \approx 1.032
\pi^2$, $\lambda_4 \approx 1.036\pi^2$ and $\lambda_5\approx
1.044\pi^2$.}
\label{fig:cross_eigen}
\end{figure}

Figure \ref{fig:cross_eigen} presents first eigenfunctions for the
crossing of two rectangular branches with $a_i = 5$ (the second
eigenfunction, which looks similar to the third one, is not shown).
As predicted by the sufficient condition (\ref{eq:cross_cond}), the
first eigenfunction (with $\lambda_1 \approx 0.66\pi^2$) is clearly
localized in the basic domain and exponentially decaying in the
branches.  In turn, all other eigenfunctions (with $\lambda_n >
\pi^2$) are not localized.

It is worth noting again that any increase of the basic domain (see
Fig. \ref{fig:cross}b) reduces the eigenvalue and thus favors the
localization.

\subsection{ Bent strip }
\label{sec:strip}

In previous examples, the basic domain was the unit square.  We
consider another shape for which the analytical estimates can be
significantly advanced.  This is a sector of the unit disk which can
be seen as a connector between two parts of a bent strip
(Fig. \ref{fig:bent}b).  In contrast to the case of infinite stripes
for which Goldstone and Jaffe have proved the existence of a localized
eigenmode for any bending (except the straigh strip)
\cite{Goldstone92}, there is a minimal branch length required for the
existence of a localized eigenmode in a finite bent strip.  In order
to demonstrate this result, we consider the family of trial functions
\begin{equation}
v_\alpha(r) = \frac{\sin \pi r}{r^\alpha} \qquad (0 < \alpha < 1) .
\end{equation}
In \ref{sec:bent}, we derive Eqs. (\ref{eq:beta_bent},
\ref{eq:sigma_bent}, \ref{eq:kappa_bent}) for the coefficients
$\beta$, $\sigma_i$, and $\kappa_i$, respectively.  Since all these
coefficients depend on $\alpha$, its variation can be used to maximize
the threshold $\eta$ given by Eq. (\ref{eq:eta}).  The numerical
computation of these coefficients suggests that $\eta$ is maximized
for $\alpha$ around $1/3$: $\eta \approx 0.7154$.  If $a_1 = a_2 = a$,
one gets the upper bound $a_{\rm th}$ of the minimal branch length
which ensures the emergence of the localized eigenfunction:
\begin{equation*}
a > a_{\rm th} = \frac{2}{\eta} \approx 2.7956 .  
\end{equation*}
We remind that this is sufficient, not necessary condition.  The
numerical computation of the first eigenvalue in the bent strip (by
FEM implemented in Matlab PDEtools) yields $a_{\rm min} \approx 2.44$.
One can see that the upper bound $a_{\rm th}$ is relatively close to
this value.  The behavior of the eigenvalue $\lambda_1$ as a function
of the branch length $a$ is shown by dashed line on
Fig. \ref{fig:lambda}.

\subsection{ Waveguide without localization }
\label{sec:triangular}

\begin{figure}
\begin{center}
\includegraphics[width=100mm]{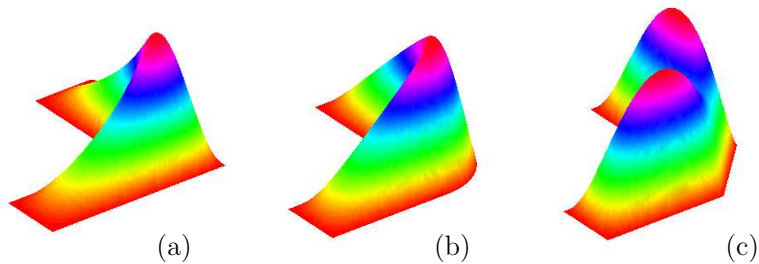}
\end{center}
\caption{
The first eigenfunction for three bent waveguides shown on
Fig. \ref{fig:bent} ($\ell = 0$), with $a = 2$.  The associate
eigenvalue $\lambda_1$ is equal to $0.9357 \pi^2$, $1.0086 \pi^2$ and
$1.1435\pi^2$, respectively.  Although the first eigenmode is
localized, all three eigenfunctions visually look similar.}
\label{fig:bent_eigenf_small}
\end{figure}

Any increase of the basic domain $\Omega$ reduces the eigenvalues and
thus preserves the localization.  In turn, a decrease of $\Omega$ may
suppress the trapped mode.  For instance, the passage from L-shape
($\Omega$ being the unit square) to the bent strip ($\Omega$ being the
quarter of the disk) led to larger minimal length required for keeping
the localization ($a_{\rm min}\approx 2.44$ instead of $a_{\rm min}
\approx 0.84$).  For instance, when $a_1 = a_2 = 2$, the first
eigenfunction, which was localized in the L-shape, is not localized in
the bent strip (Fig. \ref{fig:bent_eigenf_small}).  Further decrease
of the basic domain $\Omega$ may completely suppress the localization.

In order to illustrate this point, we consider the truncated L-shape
shown on Fig. \ref{fig:bent}c with $\ell = 0$ for which
\begin{equation*}
\Omega = \{ (x,y)\in\R^2~:~ -1 < x < 0,~ -1 < y <0, ~ x+y > - 1 \} 
\end{equation*}
is a triangle.  It is easy to see that $u(x,y) = \cos(\pi x) +
\cos(\pi y)$ is the first eigenfunction of the following eigenvalue
problem in $\Omega$:
\begin{equation*}
 - \Delta u = \tilde{\mu} u  \quad \mathrm{in}~\Omega, \qquad u|_{\Gamma} = 0,  \qquad \frac{\dc u}{\dc n}|_{\Gamma_i} = 0 ,
\end{equation*}
with the eigenvalue $\tilde{\mu}_1 =\pi^2$.  From the variational
principle
\begin{equation*}
\tilde{\mu}_1 = \inf \limits_{v \in H^1_0(\Omega), v \ne 0} 
\frac{\bigl(\nabla v, \nabla v\bigr)_{L_2(\Omega)}}{\bigl(v, v\bigr)_{L_2(\Omega)}} ,
\end{equation*}
so that 
\begin{equation*}
\bigl(\nabla v, \nabla v\bigr)_{L_2(\Omega)} \geq \tilde{\mu}_1 \bigl(v, v\bigr)_{L_2(\Omega)} 
= \pi^2 \bigl(v, v\bigr)_{L_2(\Omega)} \quad \forall v \in H^1_0(\Omega) .
\end{equation*}
Moreover, the Friedrichs-Poincar\'e inequality in the branches $Q_i$
implies \cite{Lions}
\begin{equation*}
\bigl(\nabla v, \nabla v\bigr)_{L_2(Q_i)} \geq \pi^2 \bigl(v,v\bigr)_{L_2(Q_i)} \quad \forall v \in H^1_0(Q_i) ,
\end{equation*}
from which
\begin{equation*}
\bigl(\nabla v, \nabla v\bigr)_{L_2(D)} \geq \pi^2 \bigl(v,v\bigr)_{L_2(D)} \quad \forall v \in H^1_0(D) .
\end{equation*}
As a consequence, all eigenvalues of the original eigenvalue problem
(\ref{1}) in $D$ exceed $\pi^2$ and the corresponding eigenfunctions
are not localized in the basic domain $\Omega$, whatever the length of
the branches.

When one varies continuously $\ell$ in Fig. \ref{fig:bent}c, the basic
domain transforms from the unit square (Fig. \ref{fig:bent}a) to
triangle so that one can get any prescibed minimal length $a_{\rm
min}$ between $0.84$ and infinity.  In other words, for any prescribed
branch length, one can always design such a basic domain (such $\ell$)
for which there is no localized eigenmodes.  As a consequence, the
localization may be very sensitive to the shape of the basic domain
and to the length of branches.  These effects which were overseen for
infinite waveguides, may be important for microelectronic
applications.

\begin{figure}
\begin{center}
\includegraphics[width=100mm]{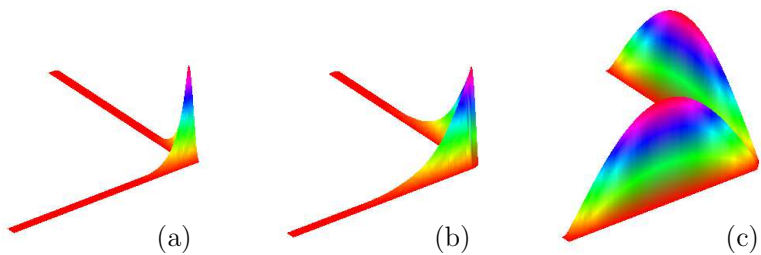}
\end{center}
\caption{
The first eigenfunction for three bent waveguides shown on
Fig. \ref{fig:bent} ($\ell = 0$), with $a = 20$.  The associate
eigenvalue $\lambda_1$ is equal to $0.9302 \pi^2$, $0.9879 \pi^2$ and
$1.0032 \pi^2$, respectively.  Although the last two values are very
close to each other, the behavior of the eigenfunctions is completely
different.}
\label{fig:bent_eigenf}
\end{figure}

Figure \ref{fig:bent_eigenf} shows the first eigenfunction for three
bent waveguides from Fig. \ref{fig:bent} with $a = 20$.  The associate
eigenvalue $\lambda_1$ is equal to $0.9302 \pi^2$, $0.9879 \pi^2$ and
$1.0032 \pi^2$, respectively.  Although the last two values are very
close to each other, the behavior of the associated eigenfunctions is
completely different.  According to the sufficient condition, the
first two eigenfunctions are localized in the basic domain, while the
last one is not.  One can clearly distinguish these behaviors on
Fig. \ref{fig:bent_eigenf}.

\subsection{ Two coupled waveguides }

\begin{figure}
\begin{center}
\includegraphics[width=130mm]{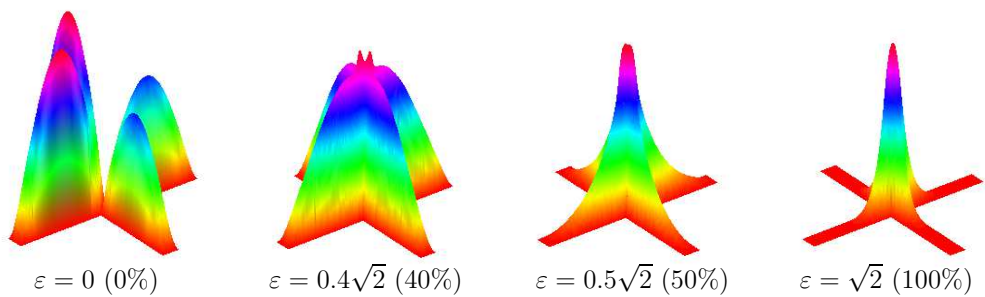}
\end{center}
\caption{
The first eigenfunction for two coupled waveguides shown on
Fig. \ref{fig:cross}c, with $a_i = 5$ and different coupling (opening
$\ve$): $\ve = 0$ (fully separated waveguides, zero coupling), $\ve =
0.4 \sqrt{2}$ (opening $40\%$ of the diagonal), $\ve = 0.5 \sqrt{2}$
(opening $50\%$ of the diagonal) and $\ve = \sqrt{2}$ (fully coupled
waveguides, no barrier).  The associate eigenvalue $\lambda_1$ is
equal to $1.05\pi^2$, $1.02\pi^2$, $0.97\pi^2$, and $0.67\pi^2$,
respectively.  In the first two cases, the eigenmodes is not
localized.  Changing the opening $\ve$, one passes from non-localized
to localized eigenmodes. }
\label{fig:cross_eigenf_eps}
\end{figure}

The coupling of infinite waveguides has been intensively investigated
\cite{Exner96}.  We consider a coupling of two finite crossing
waveguides through an opening of variable size $\ve$ as shown on
Fig. \ref{fig:cross}c.  When $\ve = 0$, one has two decoupled
waveguides from Fig. \ref{fig:bent}c for which we proved in the
previous subsection the absence of localized eigenmodes.  When $\ve =
\sqrt{2}$, there is no barrier and the waveguides are fully coupled.
This is the case of crossing between two rectangular branches as shown
on Fig. \ref{fig:cross}a, for which we checked the existence of
localized eigenmodes under weak conditions (\ref{eq:cond_L-shape}) or
(\ref{eq:cross_cond}).  Varying the opening $\ve$ from $0$ to
$\sqrt{2}$, one can continuously pass from one situation to the other.
This transition is illustrated on Fig. \ref{fig:cross_eigenf_eps}
which presents the first eigenfunction for two coupled waveguides
shown on Fig. \ref{fig:cross}c, with $a_i = 5$ and different coupling
(opening $\ve$).  The first two eigenfunctions, with $\ve = 0$ (fully
separated waveguides) and $\ve = 0.4 \sqrt{2}$ (opening $40\%$), are
not localized, while the last two eigenfunctions, with $\ve = 0.5
\sqrt{2}$ (opening $50\%$) and $\ve =
\sqrt{2}$ (fully coupled waveguides, i.e. a cross), are localized.
The critical coupling $\ve_c$, at which the transition occurs (i.e.,
for which $\lambda_1 = \pi^2$) lies between $40\%$ and $50\%$.
Although numerical computation may allow one to estimate $\ve_c$ more
accurately, we do not perform this analysis because the value $\ve_c$
anyway depends on the branch lengths.  In general, for any $a > a_{\rm
min} \approx 0.84$, there is a critical value $\ve_c(a)$ for which
$\lambda_1 = \pi^2$.  For $\ve < \ve_c$, there is no localized modes,
while for $\ve > \ve_c$ there is at least one localized mode.  The
high sensitivity of the localization character to the opening $\ve$
and to the branch lengths can potentially be employed in quantum
switching devices (see \cite{Timp88} and references therein).  

\section*{Conclusion}

We investigated the problem of trapped or localized eigenmodes of the
Laplace operator in resonators with long branches that we called
``finite waveguides''.  In this context, the localization was
understood as an exponential decay of an eigenfunction inside the
branches.  This behavior was related to the smallness of the
associated eigenvalue $\lambda$ in comparison to the first eigenvalue
of the Laplace operator in the cross-section of the branch with
Dirichlet boundary condition.  Using the explicit representation of an
eigenfunction in branches, we proposed a general variational formalism
for checking the existence of localized modes.  The main result of the
paper is the sufficient condition (\ref{eq:cond_general}) on the
branch lengths for getting a trapped mode.  In spite of the generality
of the formalism, a practical use of the sufficient condition relies
on an intuitive choice of the trial function in the basic domain
(without branches).  This function should be as close as possible to
the (unknown) eigenfunction.  Although there is no general recipe for
choosing trial functions, one can often guess an appropriate choice,
at least for relatively simple domains.

These points were illustrated for several typical waveguides,
including 2D and 3D L-shapes, crossing of the rectangular stripes, and
bent stripes.  For all these cases, the basic domain was simple enough
to guess an appropriate trial function in order to derive an explicit
sufficient condition for getting at least one localized mode.  In
particular, we obtained the upper bound of the minimal branch length
which is sufficient for localization.  We proved the existence of a
trapped mode in finite L-shape, bent strip and cross of two stripes
provided that their branches are long enough, with an accurate
estimate on the required minimal length.  These results were confirmed
by a direct numerical resolution of the original eigenvalue problem by
finite element method implemented in Matlab PDEtools.  The presented
method can be applied for studying the localization in many other
waveguides, e.g. smooth bent strip \cite{Olendski10}, sharply bent
strip \cite{Carini93} or Z-shapes \cite{Carini97}.

It is worth emphasizing that the distinction between localized and
non-localized modes is much sharper in infinite waveguides than in
finite ones.  Although by definition the localized eigenfunction in a
finite waveguide decays exponentially, the decay rate may be
arbitrarily small.  If the branch is not long enough, the localized
mode may be visually indistinguishable from a non-localized one, as
illustrated on Fig. \ref{fig:bent_eigenf_small}.  In turn, the
distinction between localized and non-localized modes in infinite
waveguides is always present, whatever the value of the decay rate.

The main practical result is an explicit construction of two families
of waveguides (truncated L-shapes on Fig. \ref{fig:bent}c and coupled
waveguides on Fig. \ref{fig:cross}c), for which the minimal branch
length $a_{\rm min}$ for getting a trapped mode continuously depends
on the parameter $\ell$ or $\ve$ of the basic domain.  For any
prescribed (long enough) branch length, one can thus construct two
almost identical finite waveguides, one with and the other without a
trapped mode.  The high sensitivity of the localization character to
the shape of the basic domain and to the length of branches may
potentially be used for switching devices in microelectronics.

\section*{Acknowledgments}

This work has been partly supported by the RFBR N 09-01-00408a grant
and the ANR grant ``SAMOVAR''.


\appendix
\section{ L-shape in three dimensions }
\label{sec:Lshape_3D}

\begin{figure}
\begin{center}
\includegraphics[width=50mm]{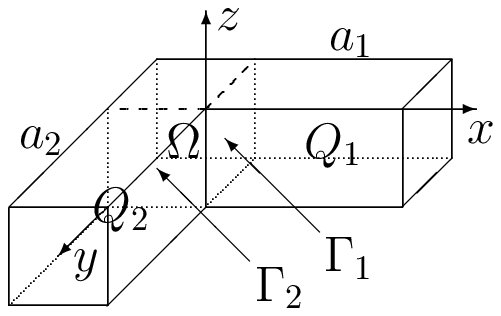}
\end{center}
\caption{
L-shape in three dimensions for which the basic domain $\Omega =
[-1,0]^3$ is the unit cube.}
\label{fig:Lshape_3D}
\end{figure}

As we mentioned at the end of Sec. \ref{sec:main}, an extension to
other types of branches is straightforward.  We illustrate this point
by considering the L-shape in three dimensions, i.e. two connected
parallelepipeds of cross-section in the form of the unit square, for
which the basic domain $\Omega$ is the unit cube
(Fig. \ref{fig:Lshape_3D}).  For each branch, the eigenvalues and
eigenfunctions in Eq. (\ref{eq:psi}) for the cross-section can be
parameterized by two indexes $m$ and $n$: $\nu_{m,n} = \pi^2 (m^2 +
n^2)$ and $\psi_{m,n}(y,z) = 2\sin(\pi m y)\sin(\pi n z)$ (similar for
the second branch).

We take the trial function
\begin{equation*}
v(x,y,z) = \bigl[(1+x)\sin(\pi y) + (1+y)\sin(\pi x)\bigr] \sin(\pi z),
\end{equation*}
which satisfies the Dirichlet boundary condition.  The coefficients
$\beta$, $\sigma_i$ and $\kappa_i$ can be found from
Eqs. (\ref{eq:beta_gen}, \ref{eq:sigma_gen}, \ref{eq:kappa_gen}), for
which the explicit integration yields
\begin{eqnarray*}
 && \bigl(v,v\bigr)_{L_2(\Omega)} = \int\limits_{-1}^0 dx \int\limits_{-1}^0 dy \int\limits_{-1}^0 dz~ v^2
= \frac{1}{6} + \frac{1}{\pi^2} , \\
&& \bigl(\nabla v,\nabla v\bigr)_{L_2(\Omega)} = \int\limits_{-1}^0 dx \int\limits_{-1}^0 dy \int\limits_{-1}^0 dz
 \biggl[\biggl(\frac{\partial v}{\partial x}\biggr)^2 + \biggl(\frac{\partial v}{\partial y}\biggr)^2 
+ \biggl(\frac{\partial v}{\partial z}\biggr)^2\biggr] = \frac{\pi^2}{3} + \frac{3}{2} , \\
\end{eqnarray*}
so that $\beta = 2\pi^2 \bigl(\nabla v,\nabla v\bigr)_{L_2(\Omega)} -
\bigl(v,v\bigr)_{L_2(\Omega)} = 1/2$.
\begin{eqnarray*}
&& \bigl(v, \psi_{1,1}\bigr)_{L_2(\Gamma_1)} = \int\limits_{-1}^0 dy \int\limits_{-1}^0 dz ~ v(0,y,z) ~ 2\sin(\pi y)\sin(\pi z) = \frac12 , \\
&& \bigl(v, \psi_{m,n}\bigr)_{L_2(\Gamma_1)} = \int\limits_{-1}^0 dy \int\limits_{-1}^0 dz ~ v(0,y,z) ~ 2\sin(\pi m y)\sin(\pi n z) = 0 
\end{eqnarray*}
(for $m \ne 1$ or $n\ne 1$), from which $\sigma_1 = \sigma_2 = 1/4$
and $\kappa_1 = \kappa_2 = 0$.  The condition (\ref{eq:cond_general})
reads as
\begin{equation*}
\frac{1}{a_1} + \frac{1}{a_2} < 2 .
\end{equation*}
If the branches have the same length, $a_1 = a_2 = a$, then the upper
bound of the minimal branch length for getting a localized
eigenfunction is given by $a_{\rm th} = 1$, as in two dimensions.

\section{ Computation for bent strip }
\label{sec:bent}

The computation of the coefficients $\beta$ and $\sigma_i$ is
straighforward, while that for $\kappa_i$ requires supplementary
estimates.

{\bf Coefficient $\beta$.} One has 
\begin{equation*}
(v,v)_{L_2(\Omega)} = \frac{\pi}{2} \int\limits_0^1 r^{1-2\alpha} \sin^2(\pi r) dr  
= \frac{\pi}{4}\biggl[\frac{1}{2(1-\alpha)} -  w_{2\alpha-1}(2\pi) \biggr],
\end{equation*}
where 
\begin{equation*}
w_\nu(q) \equiv \int\limits_0^1 r^{-\nu} \cos(q r) dr .
\end{equation*}

Similarly,
\begin{equation*}
(\nabla v_\alpha,\nabla v_\alpha)_{L_2(\Omega)} = \frac{\pi}{2} \int\limits_0^1 r (v'_\alpha)^2 dr = 
\frac{\pi}{2}\int\limits_0^1 r \biggl(\frac{\pi \cos \pi r}{r^\alpha} - \frac{\alpha \sin \pi r}{r^{1+\alpha}}\biggr)^2 dr .
\end{equation*}
Expanding the quadratic polynomial and integrating by parts, one gets
\begin{equation*}
(\nabla v_\alpha,\nabla v_\alpha)_{L_2(\Omega)} = \frac{\pi^3}{4}\biggl[\frac{1}{2(1-\alpha)} + \frac{w_{2\alpha-1}(2\pi)}{1-2\alpha} \biggr] .
\end{equation*}
Combining this term with the previous result yields
\begin{equation}
\label{eq:beta_bent}
\beta = \frac{\pi^3}{4} \frac{2\alpha}{2\alpha-1} w_{2\alpha-1}(2\pi) .
\end{equation}
In the limit $\alpha \to 1/2$, one has
\begin{equation*}
\lim\limits_{\alpha\to 1/2} \frac{w_{2\alpha-1}(2\pi)}{2\alpha-1} = \frac{\mathrm{Si}(2\pi)}{2\pi} \approx 0.2257,
\end{equation*}
where $\mathrm{Si}(x)$ is the integral sine function.

{\bf Coefficients $\sigma_i$.} For these coefficients, one gets
\begin{equation*}
\bigl(v, \sin(\pi r)\bigr)_{L_2(\Gamma_i)} = \int\limits_0^1 r^{-\alpha} \sin^2(\pi r) dr =
\frac12\biggl(\frac{1}{1-\alpha} - w_\alpha(2\pi) \biggr) ,
\end{equation*}
from which 
\begin{equation}
\label{eq:sigma_bent}
\sigma_1 = \sigma_2 = \frac12\biggl(\frac{1}{1-\alpha} - w_\alpha(2\pi) \biggr)^2 .
\end{equation}

{\bf Coefficients $\kappa_i$.} One considers 
\begin{eqnarray*}
\bigl(v, \sin(\pi nr)\bigr)_{L_2(\Gamma_i)} &=& \int\limits_0^1 r^{-\alpha} \sin (\pi r) \sin (\pi n r) dr \\
&=& \frac12 \biggl[w_\alpha(\pi(n-1))  - w_\alpha(\pi(n+1))\biggr] .
\end{eqnarray*}
The function $w_\alpha(q)$ can be decomposed into two parts
\cite{Gradsteyn},
\begin{equation*}
w_\alpha(q) = \int\limits_0^\infty r^{-\alpha} \cos(q r) dr
- \int\limits_1^\infty r^{-\alpha} \cos(q r) dr = q^{\alpha-1}  
\frac{\sqrt{\pi} ~\Gamma(\frac{1-\alpha}{2})}{2^\alpha \Gamma(\frac{\alpha}{2})} - \tilde{w}_\alpha(q) ,
\end{equation*}
where
\begin{equation*}
\tilde{w}_\alpha(q) \equiv \int\limits_1^\infty r^{-\alpha} \cos(q r) dr .
\end{equation*}

We have
\begin{equation*}
\kappa_i = 2\pi \sum\limits_{n=2}^\infty \sqrt{n^2-1} ~\bigl(v, \sin(\pi nr)\bigr)^2_{L_2(\Gamma_i)} = 
2\pi \sum\limits_{n=2}^\infty \sqrt{n^2-1}~ (d_n - e_n)^2 ,
\end{equation*}
where
\begin{eqnarray*}
d_n &=& \frac{\pi^{\alpha - \frac12}~ \Gamma(\frac{1-\alpha}{2})}{2^{1+\alpha} 
\Gamma(\frac{\alpha}{2})} \biggl[ (n-1)^{\alpha-1} - (n+1)^{\alpha-1}\biggr], \\
e_n &=& \frac12 \biggl[\tilde{w}_\alpha(\pi(n-1)) - \tilde{w}_\alpha(\pi(n+1))\biggr] .
\end{eqnarray*}

In order to estimate the coefficients $e_n$, the function
$\tilde{w}_\alpha(q)$ is integrated by parts that yields for $q =
\pi(n\pm 1)$:
\begin{eqnarray*}
\tilde{w}_\alpha(q) &=& \alpha \frac{\cos q}{q^2} - \alpha(\alpha+1) \frac{\tilde{w}_{\alpha+2}(q)}{q^2} \\
&=& \alpha \frac{\cos q}{q^2} - \alpha(\alpha+1)(\alpha+2) \frac{\cos q}{q^4} 
+ \alpha(\alpha+1)(\alpha+2)(\alpha+3) \frac{\tilde{w}_{\alpha+4}(q)}{q^4} .
\end{eqnarray*}
The inequality
\begin{equation*}
(\alpha+3)|\tilde{w}_{\alpha+4}(q)| \leq (\alpha+3) \int\limits_1^\infty r^{-\alpha-4} dr = 1
\end{equation*}
leads to
\begin{equation*}
\fl \alpha \frac{\cos q}{q^2} - \alpha(\alpha+1)(\alpha+2) \frac{\cos q + 1}{q^4} \leq
\tilde{w}_\alpha(q) \leq \alpha \frac{\cos q}{q^2} - \alpha(\alpha+1)(\alpha+2) \frac{\cos q - 1}{q^4} ,
\end{equation*}
from which
\begin{equation*}
e_n^- \leq e_n \leq e_n^+  ,
\end{equation*}
where the lower and upper bounds are
\begin{eqnarray*}
e_n^\pm &=& \frac12 \biggl[\biggl(\frac{\alpha}{\pi^2} \frac{(-1)^{n-1}}{(n-1)^2} - \frac{\alpha(\alpha+1)(\alpha+2)}{\pi^4} 
\frac{(-1)^{n-1} \mp 1}{(n-1)^4}\biggr) \\
&& - \biggl(\frac{\alpha}{\pi^2} \frac{(-1)^{n+1}}{(n+1)^2} - \frac{\alpha(\alpha+1)(\alpha+2)}{\pi^4} 
\frac{(-1)^{n+1} \pm 1}{(n+1)^4}\biggr)\biggr] .
\end{eqnarray*}
Using these estimates, one gets
\begin{eqnarray*}
\sum\limits_{n=2}^\infty \sqrt{n^2-1}~ d_n^2 &\equiv& A_1 ,\\
\sum\limits_{n=2}^\infty \sqrt{n^2-1}~ d_n e_n &\geq&  \sum\limits_{n=2}^\infty \sqrt{n^2-1}~ d_n e_n^- \equiv A_2^- ,\\
\sum\limits_{n=2}^\infty \sqrt{n^2-1}~ e_n^2   &\leq&  \sum\limits_{n=2}^\infty \sqrt{n^2-1}~ (e_n^+)^2 \equiv A_3^+ ,
\end{eqnarray*}
from which
\begin{equation}
\label{eq:kappa_bent}
\kappa_i \leq \kappa, \qquad  \kappa \equiv 2\pi (A_1 - 2A_2^- + A_3^+) .
\end{equation}
Although the expressions for $A_1$, $A_2^-$ and $A_3^+$ are
cumbersome, the convergence of these series can be easily checked,
while their numerical evaluation is straightforward.

The numerical computation of these coefficients shows that the
threshold value $\eta$ is maximized at $\alpha \approx 1/3$: $\eta
\approx 0.7154$.  Note that if the coefficients $\kappa_i$ were
computed by direct numerical integration and summation, the value of
$\eta$ for $\alpha = 1/3$ could be slightly improved to be $0.7256$.
The difference results from the estimates we used, and its smallness
indicates that the estimates are quite accurate.

\end{document}